\documentclass[twocolumn,superscriptaddress, showpacs,preprintnumbers,pra]{revtex4-1}
\usepackage{dcolumn}
\usepackage{graphicx}
\usepackage{amsmath, amssymb}
\usepackage{bm}
\usepackage{mathrsfs}
\usepackage{subfigure}
\usepackage{dsfont}
\usepackage{latexsym}
\usepackage{amsthm}
\usepackage{subfigure}
\usepackage{dsfont}
\usepackage{color}

\newcommand{\vecc}[1]{\mbox{\boldmath $#1$}}

\DeclareGraphicsExtensions{.pdf,.ps,.eps}

\begin{document}

\title{Ferromagnetic ground state of the SU($3$) Hubbard model on the Lieb lattice}

\author{Wenxing Nie}
\affiliation{College of Physical Science and Technology, Sichuan University, Chengdu, Sichuan 610064, China}
\author{Deping Zhang}
\affiliation{Institute for Advanced Study, Tsinghua University, Beijing, 100084, China}
\author{Wei Zhang}
\thanks{Corresponding author: {wzhangl@ruc.edu.cn}}
\affiliation{Department of Physics, Renmin University of China, Beijing 100872, China}
\affiliation{Beijing Key Laboratory of Opto-electronic Functional Materials and Micro-nano Devices,
Renmin University of China, Beijing 100872, China}

\date{\today}

\begin{abstract}
We investigate the magnetic properties of a repulsive fermionic SU($3$) Hubbard model on the Lieb lattice from weak to strong interaction by means of the mean-field approximation. To validate the method we employed, we first discuss the SU($2$) Hubbard model at the mean-field level, and find that our results are consistent with known rigorous theorems. We then extend the calculation to the case of SU($3$) symmetry. We find that, at $4/9$ filling, the SU$(3)$ symmetry spontaneously breaks into the SU$(2)\times$U$(1)$ symmetry in the ground state, leading to a staggered ferromagnetic state for any repulsive $U$ at zero temperature. 
We then investigate the stability of the ferromagnetic state by relaxing the filling away from $4/9$, and conclude that the ferromagnetic state is sensitive but robust to fillings, as it can persist within a certain filling regime.
We also apply the mean-field approximation to finite temperature to calculate the critical temperature and the critical entropy of the ferromagnetic state. As the resulting critical entropy per particle is significantly greater than that can be realized in experiments, we expect some quasi-long-range-ordered features of such a ferromagnetic state can be realized and observed with fermionic alkaline-earth-metal(-like) atoms loaded into optical lattices.
\end{abstract} 

\date{\today}
\maketitle

\section{\label{intro}Introduction}

The investigation of itinerant ferromagnetism is one of the fundamental and central topics in theoretical condensed matter physics~\cite{Lieb-Mattis,Nagaoka, Mielke1,Mielke2,TasakiPRL,Mielke-Tasaki, Lieb-theorem, YiLi,Wu-QMC-thermo-PRX2015,HongYao_Nagaoka2D,Wu-Sp4Mag-PRB2011}. Although itinerant ferromagnetism is generally considered as a result of strong electron-electron interaction since paramagnetism is an inevitable consequence of the Pauli exclusion principle in non-interacting many-electron systems, strong interaction does not necessarily lead to ferromagnetism. For instance, Lieb and Mattis proved that ferromagnetism never happens in one-dimensional systems with only nearest neighbor hopping, no matter how strong the interaction is~\cite{Lieb-Mattis}. Up to date, the stability of ferromagnetism in itinerant fermions remains a challenging problem in condensed matter physics, mainly because of the subtle interplay between the kinetic energy and the interaction energy. 

Due to the subtle sign problem, rigorous proofs for ferromagnetism of the fermionic Hubbard model are very rare~\cite{Nagaoka, Mielke1,Mielke2,TasakiPRL,Mielke-Tasaki, Lieb-theorem,YiLi}. The existing fews hence play important roles as benchmarks for the magnetic properties of the ground state. One of the earliest rigorous proofs of itinerant ferromagnetism is attributed to Thouless~\cite{Thouless} and Nagaoka~\cite{Nagaoka}. It is shown that, with \emph{infinite} on-site repulsion and single hole away from half filling,  the ground state of the Hubbard model defined on a lattice satisfying connectivity condition is unique with saturated ferromagnetism (with maximal total spin $S=N_e/2$, where $N_e$ is the number of electrons). In 1989, Lieb proved that on a general bipartite lattice where the two consisting sublattices have different numbers of sites, the ground state of the Hubbard model at half filling with any \emph{finite} repulsion is an unsaturated ferromagnetic state, which is also referred as a ferrimagnetic state with total spin being proportional to system size, $S=\big{|} |\Lambda_{ \textrm{A}}| - |\Lambda_{\textrm{B}}| \big{|}/2$. Here, $|\Lambda_{ \textrm{A(B)}}|$ is the number of sites on sublattice A (B)~\cite{Lieb-theorem}. In such a setting, a consequence of the bipartite condition is the existence of a non-dispersive, or equivalently, a flat band within the single-electron spectrum. Such a non-dispersive band implies the existence of localized single-particle eigenstates in spatial space and degeneracy of eigenstates in the single-particle spectrum. This type of ferromagnetism is then referred as flat-band ferromagnetism, and is further proved by Mielke and Tasaki~\cite{Mielke1,Mielke2,TasakiPRL,Mielke-Tasaki}. The role of the flat band can be understood in the following argument. In a non-interacting system, the ferromagnetic state is one of the many degenerate ground states when particles fill into the flat band. As one turns on the interaction between electrons, such degeneracy is lifted by perturbation and the ferromagnetic ground state may be singled out to be the only ground state. Although such an argument only holds in the non- and weak-interacting limit and completely breaks down in strongly interacting many-electron systems, a lattice featuring a flat band may still be beneficial to realize ferromagnetism as it effectively reduces kinetic energy. 

These two seemingly different approaches to realize ferromagnetism, the Nagaoka's ferromagnetism~\cite{Nagaoka} and the flat-band ferromagnetism~\cite{Mielke1,Mielke2,TasakiPRL,Mielke-Tasaki,Lieb-theorem}, can be understood in a unified way in the context of the Stoner's criterion for ferromagnetism: $\rho(E_{\textrm{F}}) U >1$. Here, $\rho(E_{\textrm{F}})$ is the density of states (DOS) at the Fermi level and $U$ is the inter-particle Coulomb interaction. Although this argument is obtained from a mean-field calculation, it provides us the picture that there may be an instability towards ferromagnetism when the DOS at the Fermi level and/or the Coulomb interaction become sufficiently large. The two theorems mentioned above could be thought as satisfying the condition of Stoner's criterion: either through infinite interaction, or through infinite DOS at the Fermi level. 

As the discussion above focus on fermionic systems possessing the SU$(2)$ spin-rotational symmetry, one would wonder what is the ground state of the Hubbard model which acquires a higher SU$(N)$ spin-rotational symmetry. The SU$(N)$ symmetry was introduced in condensed matter physics as a mathematical extension or theoretical tool for strongly correlated SU$(2)$ systems\cite{Affleck1,Affleck2,Sachdev1,Sachdev2}. 
However, the degrees of freedom offered by the nuclear spin of cold atoms offer an fascinating playground to investigate the strongly correlated particles with a higher symmetry. The simulation of the SU$(N)$ Hubbard model in cold atoms has not only been theoretically proposed~\cite{SO5-WuPRL2003,Wu-review,Honerkamp1,Honerkamp2,Sr-proposal}, but also been experimentally realized by loading ultracold alkaline-earth-metal(-like) atoms in various types of optical lattices~\cite{Sr-Rice, Bloch, Fallani, Takahashi_PRL, Takahashi_Nat}. Because of the decoupling of electronic and nuclear spin degrees of freedom, the alkaline-earth-metal(-like) atoms with different nuclear spin magnetic quantum numbers have an SU$(N)$ symmetry, where $N=2I+1$ is the number of Zeeman states, and $I$ is the nuclear spin. The number $N$ can be as large as 10 for $^{87}$Sr with $I=9/2$, or 6 for $^{173}$Yb with $I=5/2$. 

The SU($N>2$) generalization of the itinerant ferromagnetism has been previously studied with an SU$(N)$ Hubbard model~\cite{Katsura, Cazalilla}. In Ref.~\cite{Katsura}, a sufficient condition is provided for an extension of the Nagaoka state by a rigorous proof. In Ref.~\cite{Cazalilla}, the authors determine the transition from paramagnetic to itinerant ferromagnetic states with increasing scattering length, and derive the SU($N$) generalization of the Stoner's criterion for ferromagnetism at the mean-field level, finding that it is the same as the SU($2$) case. As the Nagaoka's ferromagnetism~\cite{Nagaoka} and the flat-band ferromagnetism~\cite{Mielke1,Mielke2,TasakiPRL,Mielke-Tasaki,Lieb-theorem} in SU($2$) systems could be regarded as satisfying the SU($2$) Stoner's criterion, we expect that the generalized SU($N$) ferromagnetism could be stabilized in systems satisfying the SU($N$) Stoner's criterion, for example, a system with a partially filled flat band, where the DOS diverges at the Fermi level.   

In this paper, we investigate the magnetic properties of the ground state of the SU$(3)$ Hubbard model with finite repulsion on the Lieb lattice, by means of a mean-field calculation. The Lieb lattice has many advantages to investigate this problem. First, with only nearest-neighbor hoppings, the Lieb lattice is bipartite and features a flat band in the middle of single-electron spectrum. The system acquires a diverging DOS as the Fermi level is filled into the flab band. Second, the Lieb lattice is a special lattice structure described by Lieb's theorem, so that we can compare our calculation of the SU($2$) Hubbard model with rigorous theorem. Last but not least, the Lieb lattice has been successfully realized for $^{173}$Yb atoms~\cite{Taie}, paving the way for the experimental investigation of flat-band ferromagnetism with the SU$(N)$ symmetry. Although the mean-field theory is not as reliable as rigorous theorems, it does provide us insights about the possible ferromagnetic ground state, the effect of interaction, and the temperature regime where the quasi-long-range magnetic order persists.  

The remainder of the paper is organized as follows. We first briefly review the band structure of the tight-binding model on the Lieb lattice in Sec.~\ref{lattice}. In Sec.~\ref{model}, we introduce the model with the SU($N$) symmetry and discuss the symmetries implied by the Hamiltonian. In Sec.~\ref{general-MF}, we decouple the interaction in the spin (or flavor) channel and develop a general mean-field approach for the SU$(N)$ Hubbard model. Then we discuss the simplifications and implications of the model if the ground state supports collinear magnetic orders. In Sec.~\ref{SU2}, we present the results of mean-field calculation for the SU$(2)$ Hubbard model where the ferromagnetic ground state is found, and compare our results with known rigorous theorems. In Sec.~\ref{SU3}, we present the results for the SU$(3)$ Hubbard model at $4/9$ filling. We find that the ground state is ferromagnetic, as a consequence of the spontaneous symmetry broken from SU$(3)$ to SU$(2)\times$U$(1)$. We then extend the discussion to other filling factors and find that the ferromagnetic state can persist within a certain filling regime. Furthermore, within the mean-field approximation, we find the ferromagnetic state is stable up to a finite temperature. In Sec.~\ref{conclusion}, we present the conclusions and discussions. The generators of SU$(3)$ group, i.e., the Gell-Mann matrices, and the technical part of the mean-field calculation involved herein are presented in the Appendices. 

\section{\label{lattice}Tight-binding model on the Lieb lattice}
\begin{figure}  
\centering
\includegraphics[width=0.9\columnwidth]{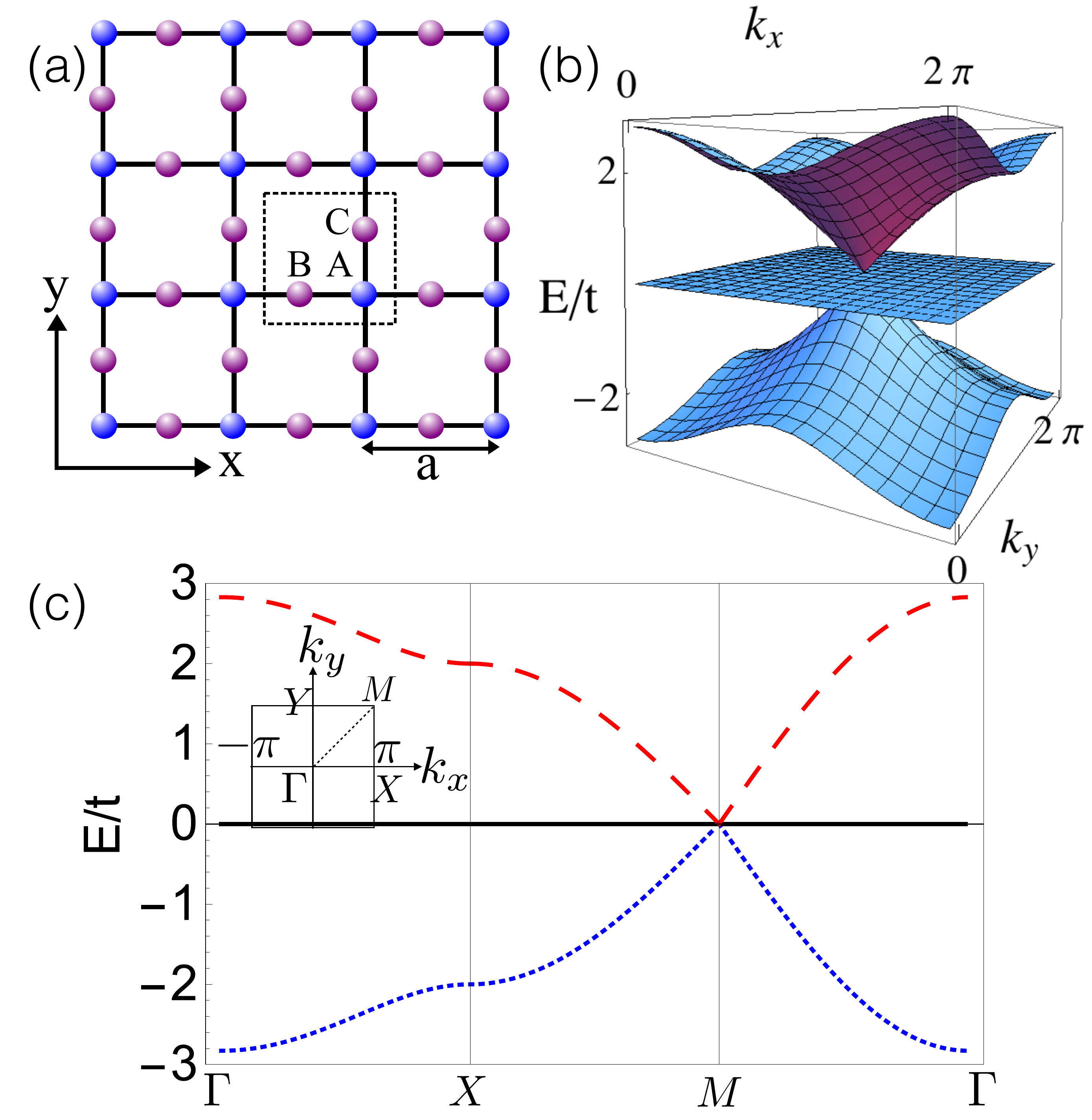}
\caption{(a) Lattice structure of the Lieb lattice. The dotted square shows one unit cell, containing three lattice sites with indices A, B, and C.  (b) The band structure of tight-binding model on the Lieb lattice. (c) The band structure along high symmetric lines in the first Brillouin Zone (BZ), with the inset shows the first BZ and high symmetric points.}
\label{fig.lattice-BZ}
\end{figure}

The two-dimensional Lieb lattice is an important building block of layered perovskite material, e.g., the CuO$_2$ plane in high-$T_c$ cuprate superconductors.
It consists of two sublattices as shown in Fig.~\ref{fig.lattice-BZ}(a): one has lattice spacing $a$ (depicted in blue), while the other one has spacing $a/\sqrt{2}$ and is tilted for $45^{\circ}$ (purple). The dotted square in Fig.~\ref{fig.lattice-BZ}(a) shows one unit cell, which contains three sites labeled by A, B and C. Notice that the site B and C may have different magnetization orders when the $C_4$ rotational symmetry is broken. This can be realized either by explicitly imposing a chemical potential imbalance between sites B and C such that the three-fold degeneracy at the$(\pi,\pi)$ point in momentum space is lifted, or by interaction effect which leads to a nematic phase with spontaneously broken $C_4$ symmetry~\cite{Zhang,Tsai}. The basis vectors of the unit cell can be chosen as $\bold{a}_1=(a,0)$ and $\bold{a}_2=(0,a)$. 

The tight-binding model for spinless fermions reads
\begin{equation}
\mathcal H_{\textrm{tb}}=-\sum_{\langle i,j\rangle\in\Lambda}  (t_{ij}c^{\dag}_{i}c_{j}+\mbox{H.c.}),
\end{equation}
where $\langle i, j\rangle$ runs over all pairs of nearest-neighbor sites on lattice $\Lambda$, $c_i^{\dag}$  $(c_i)$ is the creation (annihilation) operator for fermion on site $i$, satisfying anti-commutation relations $\{ c_i^{\dag}, c_j\}=\delta_{ij}$ and $\{ c_i, c_j\}=0$, and H.c. stands for Hermitian conjugate. For simplicity, we take the uniform real hopping amplitudes $t_{ij}=t$. With only nearest-neighbor hoppings, the Lieb lattice is bipartite and consequently features a flat band. The band structure is easily obtained by transforming $\mathcal H_{\textrm{tb}}$ into momentum space, $\mathcal H_{\textrm{tb}}=\sum_{\bold{k}} \Phi_{\bold{k}}^{\dag} H_{\textrm{tb}}(\bold{k}) \Phi_{\bold{k}}$, where $\Phi_{\bold{k}}$ is a $3$-component spinor $(c_{\textrm{A}}(\bold{k}), c_{\textrm{B}}(\bold{k}),c_{\textrm{C}}(\bold{k}))^{T}$ with sublattice indices A, B and C. The vector $\bold{k}=(k_x,k_y)$ belongs to the first Brillouin zone (BZ), which is shown in the inset of Fig.~\ref{fig.lattice-BZ}(c). $H_{\textrm{tb}}(\bold{k})$ takes the form
\begin{displaymath}
H_{\textrm{tb}}(\bold{k}) =-2t
\left( \begin{array}{ccc}
0 & \cos(k_x a/2) & \cos(k_y a /2) \\
\cos (k_x a/2) & 0 & 0\\
\cos (k_y a/2) & 0 & 0\\
\end{array}
\right).
\end{displaymath}
It has two dispersive bands $\epsilon_{\pm} (\bold{k})  = \pm t\sqrt{4+2\cos(\bold{k} \cdot \bold{a}_1)+2\cos(\bold{k} \cdot \bold{a}_2)}$, and one flat band in the middle of the spectrum $\epsilon_0 (\bold{k})  =  0$.
The three bands touch each other at the $M$ point ($\pi,\pi$) as shown in Figs.~\ref{fig.lattice-BZ}(b) and \ref{fig.lattice-BZ}(c).

The magnetic and topological properties of the Lieb lattice have been comprehensively investigated~\cite{Goldman,Cazalilla,Tsai, Zhang,Taie, HuWang,Noda_2d,Noda_layered,Noda_3d,TMI_MarDel}.
In experiment, the Lieb lattice can be realized by imposing three types of optical lattices~\cite{lattice-proposal1,lattice-proposal2,Goldman, Taie}, leading to the following potential profile: $V(x,y)=V_0[\sin^2(\pi x/a)+\sin^2(\pi y/a)]+V_1[\sin^2(2 \pi x/a)+\sin^2(2 \pi y/a)]+V_2[\cos^2(\pi x/2 a+\pi y/2 a)+\cos^2(\pi x/2 a-\pi y/2 a)]$, where $V_0=V_1=2V_2$ are the depths of composing lattice potentials.

\section{\label{model}the SU($N$) Hubbard model on the Lieb lattice}
In the following, we consider the Hubbard model with SU($N$) spin-rotational symmetry, with a \emph{finite} repulsive on-site density-density interaction $U>0$. In general, the Hamiltonian can be written as:
\begin{equation}
\mathcal H=-t \sum_{\langle i,j\rangle \in \Lambda}\sum_{\alpha=1}^N (c^{\dag}_{i,\alpha}c_{j,\alpha}+\mbox{H.c.})
+U\sum_i n_i^2,
\label{eq.Ham}
\end{equation}
where ${\alpha} = 1, \cdots, N$ labels the flavor or color of a fermion. In cold atomic systems, $N$ is the number of 
populated Zeeman states.
The operator $c_{i,\alpha}$ and its adjoint $c_{i,\alpha}^{\dag}$ satisfy the fermionic anticommutation relations $\{ c_{i,\alpha}^{\dag}, c_{j,\beta}\}=\delta_{ij}\delta_{\alpha\beta}$ and $\{ c_{i,\alpha}, c_{j,\beta}\}=0$.  The number operator for fermions at site $i$ is defined as $n_{i}=\sum_{\alpha=1}^N n_{i,\alpha}$ with $n_{i,\alpha}=c_{i,\alpha}^{\dag}c_{i,\alpha}$.

The symmetry of the Hamiltonian Eq.~\eqref{eq.Ham} has been discussed in detail in Ref.~\cite{Katsura}. Besides the trivial conservation of total particle number, the Hamiltonian exhibits a global $\textrm{U}(N)=\textrm{U}(1)\times \textrm{SU}(N)$ symmetry. One can construct a total number operator and spin operators, which commute with the Hamiltonian Eq.~\eqref{eq.Ham}:
\begin{eqnarray}
\hat{N} & = & \sum_{i\in\Lambda} n_i,\\
\hat{S}_i^a & = & \sum_{\alpha,\beta=1}^{N} c^{\dag}_{i\alpha} T^a_{\alpha\beta}c_{i\beta} \qquad (a=1,\cdots,N^2-1),\label{eq.spin-opt}
\end{eqnarray}
where $T^a=\tau^a/2$ are the generators of the SU$(N)$ Lie algebra. To be specific, for fermions with the SU$(2)$ symmetry, $\alpha=1,2$  (or equivalently $\alpha=\uparrow, \downarrow)$, and $\tau^a=\sigma^a \; (a=1,2,3)$, where 
$\sigma^a$'s are the Pauli matrices. For fermions with the SU$(3)$ symmetry, $\alpha= 1,2,3$ and $\tau^a=\lambda^a \; (a=1,\cdots,8)$, where $\lambda^a$'s are the Gell-Mann matrices (shown in Appendix A).  

For the case of the SU$(2)$ symmetry, Lieb has proved rigorously that at half filling, the ground state of Eq.~\eqref{eq.Ham} is unique (apart from the trivial $(2S+1)$-fold degeneracy) and has total spin $S=\mathscr{N}/2$, where $\mathscr{N}$ is the number of unit cells~\cite{Lieb-theorem}. Strictly speaking, it is a ferrimagnetic state because the total spin is not saturated. In Lieb's terminology, ferromagnetism means the total spin is nonzero and proportional to the systems size. In the following discussion, we adopt the same definition and do not distinguish ferrimagnetism from ferromagnetism. 

There is no general theorem for the ground state of the Hamiltonian Eq.~\eqref{eq.Ham} with SU($N>2$) symmetry with finite interaction $U$, as far as we know. For the SU($3$) Hubbard model with attractive interaction $U<0$, the spin reflection positivity (the method employed in Lieb's proof for the SU($2$) Hubbard model) can be extended with the aid of Majorana fermion representation~\cite{SRP-majorana, Zixiang-priv}. However, for repulsive interaction $U>0$, such an extension becomes problematic. Therefore, we use the Hartree-Fock mean-field approximation to study the magnetic properties of the Hubbard model with SU$(2)$ and SU$(3)$ symmetries, respectively.

\section{\label{general-MF}Mean-field approach for the SU$(N)$ Hubbard model}

Before focusing on the specific cases of the SU($2$) and SU($3$) symmetries, we present a general mean-field formalism for the SU($N$) Hubbard model, then discuss the simplifications and implications of the general form in a special scenario, when the magnetization vectors are found to be collinear. 

With the completeness relation of the generators of the SU($N$) Lie algebra, the density-density interaction can be decomposed in the spin channel (more details in Appendix B):
\begin{equation}
U\sum_i n_i^2 = -\frac{2NU}{N+1}\sum_i \bold {S}_i^2 + NU\sum_i n_i,
\label{eq.interaction}
\end{equation} 
where the SU($N$) ``spin'' operator is constructed in Eq.~\eqref{eq.spin-opt}.
Perform the mean-field approximation $\bold{S}_i^2  \simeq 2 \langle \bold{S}_i \rangle \cdot \bold{S}_i -\langle \bold{S}_i \rangle ^2$, and define the magnetization $\bold{m}_i =\langle \bold{S}_i \rangle$,  
the mean-field Hamiltonian is expressed as 
\begin{eqnarray}
\mathcal H_{\textrm{MF}} & = & -t \sum_{\langle i,j\rangle \in \Lambda}\sum_{\alpha=1}^N (c^{\dag}_{i\alpha}c_{j\alpha}+\mbox{H.c.})+\frac{2NU}{N+1}\sum_{i\in\Lambda} \bold{m}_i^2{}\nonumber\\
&&{}-\frac{4NU}{N+1} \sum_{i\in\Lambda}\sum_{\alpha,\beta=1}^N c^{\dag}_{i\alpha}\bold{m}_{i} \cdot \frac{\vecc{\tau}_{\alpha\beta}}{2}c_{i\beta}. 
\label{eq.MF-Ham}
\end{eqnarray}
Assume the translational symmetry is not broken so that the magnetization order is uniform, we can preform a Fourier transformation
$c_{\mu,\alpha} (\bold{k}) = (1/\sqrt{\mathscr{N}})\sum_{\bold{R}} e^{-i \bold{k} \cdot \bold{R}} c_{\mu,\alpha}(\bold{R})$, 
where $\bold{R}$ is the position of a unit cell, $\mu=\textrm{A,B,C}$ labels the sublattices, and $\mathscr{N}$ is the total number of unit cells. As a result, we obtain the mean-field Hamiltonian in the momentum space, 
\begin{equation}
\mathcal H_{\textrm{MF}}=\sum_{\bold{k}}\Phi^{\dag}_{\bold{k}} H_{\bold{k}}^{\textrm{MF}}\Phi_{\bold{k}}
+\frac{2NU}{N+1}\sum_i \bold{m}_i^2,
\label{eq.mf-Hmat}
\end{equation}
where $\Phi_{\bold{k}} = (c_{\textrm{A},\alpha} (\bold{k}),c_{\textrm{B},\alpha} (\bold{k}), c_{\textrm{C},\alpha} (\bold{k}))^T$ is a $3N$-component spinor with $\alpha=1,\cdots,N$,
and 
\begin{eqnarray}
H_{\bold{k}}^{\textrm{MF}} = -
\left( \begin{array}{ccc}
\frac{2NU}{N+1} \bold{m}_{\textrm{A}}\cdot \vecc{\tau} & t\cos(\frac{k_x a}{2}) \mathds{1} & t\cos (\frac{k_y a}{2})\mathds{1} \\
t\cos (\frac{k_x a}{2}) \mathds{1} &\frac{2NU}{N+1}  \bold{m}_{\textrm{B}}\cdot \vecc{\tau} & \bold{0}\\
t\cos (\frac{k_y a}{2}) \mathds{1} & \bold{0} & \frac{2NU}{N+1}  \bold{m}_{\textrm{C}}\cdot \vecc{\tau}\\
\end{array}
\right).
\label{eq.matHk}
\end{eqnarray}

The magnetization $\bold{m}_{\mu}$ is an $(N^2-1)$-dimensional vector in the spin/flavor space, which transforms according to the adjoint representation of the SU($N$) Lie algebra.
The standard mean-field self-consistent calculation goes as follows: start with an arbitrary combination of $3(N^2-1)$ parameters of $\bold{m}_{\mu}$, numerically calculate the ground state, obtain the expectation value of the spin operators $\langle \bold{S}_{\mu} \rangle$ (more details in Appendix B), and iterate until the magnetization $\bold{m}_{\mu}$ at each site converges. Note that this algorithm is equivalent to minimize the ground-state energy functional $E (\bold{m}_{\mu})=\langle \Psi(\bold{m}_{\mu}) | \mathcal H_{\textrm{MF}}| \Psi (\bold{m}_{\mu})\rangle/\langle \Psi(\bold{m}_{\mu}) | \Psi(\bold{m}_{\mu}) \rangle$ variationally.

Next, we consider a special case of ``collinear" magnetic state where the magnetization vectors on the three sublattices are ``parallel'' or ``antiparallel'' to each other. By defining the angle between the two vectors $\bold{m}_{\mu}$ and $\bold{m}_{\nu}$ in the $(N^2-1)$-dimensional spin/flavor space
$\theta_{<\bold{m}_{\mu},\bold{m}_{\nu}>}=\arccos \frac{\bold{m}_{\mu}\cdot \bold{m}_{\nu}}{|\bold{m}_{\mu}| |\bold{m}_{\nu}|}$,
where the inner product and the norm are defined in $(N^2-1)$-dimensional generalized Euclidean space: $\bold{m}_{\mu}\cdot \bold{m}_{\nu} =\sum_{a=1}^{N^2-1} m_{\mu}^a m_{\nu}^a$, and $|\bold{m}_{\mu}| =\sqrt{\bold{m}_{\mu}\cdot \bold{m}_{\mu}}$, we note that the two vectors $\bold{m}_{\mu}$ and $\bold{m}_{\nu}$ are collinear if $\cos \theta_{<\bold{m}_{\mu},\bold{m}_{\nu}>}=\pm 1$.

In such a special case, we can perform a \emph{global} SU($N$) unitary transformation to Eq.~\eqref{eq.MF-Ham} to make some simplifications,   
\begin{equation}
  \mathbf{U}\mathbf{m}_{\mu} \cdot \vecc{\tau}\mathbf{U}^{\dagger}=\sum_{a=1}^{N-1} m'^{a}\mathbf{H}^a,
  \label{eq.SUtrans}
\end{equation}
where $\mathbf{H}^a$ are the diagonal and commuting generators in the SU($N$) algebra and constitute the Cartan subalgebra of the SU($N$) Lie algebra~\cite{Georgi-book}: $\{\mathbf{H}^a,a=1,2,\cdots,N-1\}$.
Note that the Cartan subalgebra for the SU($2$) algebra is $\{\mathbf{H}^1=T^3\}$, and for the SU($3$) algebra is $\{\mathbf{H}^1=T^3,\mathbf{H}^2=T^8\}$.
Equation~\eqref{eq.SUtrans} can be understood by noting that the $N$-dimensional Hermitian matrix $\bold{m}_{\mu} \cdot \vecc{\tau}$ can be diagonalized by an SU($N$) transformation $\mathbf{U}$. Since $\bold{m}_{\mu} \cdot \vecc{\tau}$ is traceless and this property is not changed by unitary transformations, the diagonalized matrix is also traceless. As a result, there are only $N-1$ degrees of freedom in the diagonalized matrix and it can be expanded by the $N-1$ elements in Cartan subalgebra. 

This important observation can not only significantly reduce the computational time, but also imply that there should be $N$ degenerate equivalent mean-field ground states if the ground state supports collinear magnetic orders. These $N$ equivalent mean-field states are determined by the $N$ common eigenstates of the $N-1$ commuting generators in the Cartan subalgebra. As these mean-field states are related to each other through an SU$(N)$ rotations, one can arbitrarily choose one to study the magnetic properties. These equivalent state can be understood with the aid of weights diagram in $\{\mathbf{H}^a,a=1,2,\cdots,N-1\}$ plane (see detailed discussions in  Sec.~\ref{SU3}). Take our familiar SU$(2)$ models as an example: with SU($2$) symmetry spontaneous breaking, the two equivalent ground states with collinear magnetic orders can be chosen as the two eigenstates of $\sigma_z$, i.e., either spin-up polarized or spin-down polarized. It is easy to see that Eq.~(\ref{eq.SUtrans}) 
is an application of the frequently used convention in SU($2$) systems into the SU($N$) case.
We emphasize that for a system with non-collinear (e.g., chiral or spiral) magnetization orders, we could not do this simplification because one can not find a \emph{global} transformation that satisfies Eq.~\eqref{eq.SUtrans}.    

\section{\label{SU2}Mean-field calculation for the SU$(2)$ Hubbard model}

It is straightforward to perform the mean-field calculation for the SU$(2)$ Hubbard model, based on the general formalism derived for the SU$(N)$ Hubbard model. We simply need  to take $N=2$ and $\vecc{\tau}=\vecc{\sigma}$ in Eq.~\eqref{eq.matHk} and perform the self-consistent iterations as specified in the previous section.  

\begin{figure}
\centering 
\includegraphics[width=1.0\columnwidth]{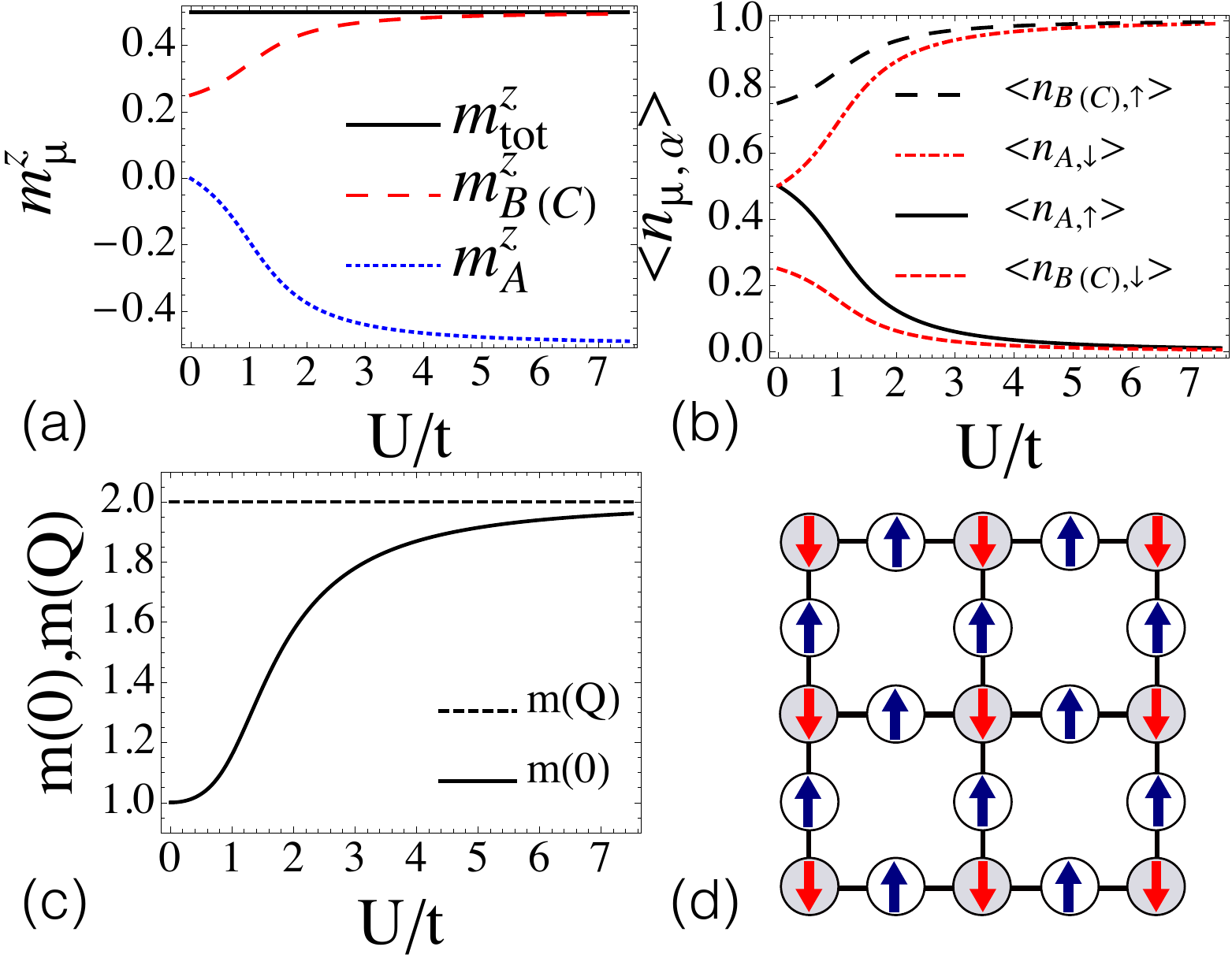}
\caption{The properties of the magnetic order of the SU(2) Hubbard model at half filling. (a) The magnetization orders on each site $m_{\mu}^z$ and total magnetization in one unit cell $m_{\textrm{tot}}^z$ as functions of $U/t$, where  $\mu=\textrm{A,B,C}$. The result of $m_{\mu}^{x,y} = 0$ is not shown. (b) The particle numbers of each spin component on each site $\langle n_{\mu,\alpha} \rangle$ as functions of $U/t$, where $\alpha=\uparrow,\downarrow$. (c) The correlation functions $m(0)$ and $m(Q)$ as functions of $U/t$. (d) A schematic figure showing the staggered ferromagnetic long-range order at large-$U$ limit.}
\label{fig.SU2}
\end{figure} 

To verify how well the mean-filed calculation works, we first investigate the magnetization order $\bold{m}_{\mu}$ at half-filling (filling fraction $\nu=1/2$, which is defined as $\nu=N_e/(2N_s)$ for the SU($2$) system with $N_e$ the number of particles and $N_s$ the number of sites). By minimizing the ground state energy functional, we find that the ground state always has a collinear magnetic order with $\bold{m}_{\textrm{B}}$ being parallel to $\bold{m}_{\textrm{C}}$, and antiparallel to $\bold{m}_{\textrm{A}}$. Besides, $\bold{m}_{\textrm{B}}=\bold{m}_{\textrm{C}}$ holds for any value of $U/t$, implying the $C_4$ symmetry is preserved. Since the magnetic order is collinear, we can perform a global SU($2$) transformation, such that only the $m^z$ component is nonzero in the new axes in spin space. 

In  Fig.~\ref{fig.SU2}(a), we plot $m_{\textrm{A}}^z $, $ m_{\textrm{B(C)}}^z $ and the total magnetization in one unit cell $m^z_{\textrm{tot}} $ with increasing $U/t$. Here, the total magnetization in one unit cell is defined as $\bold{m}_{\textrm{tot}}=\sum_{\mu=\textrm{A}, \textrm{B},\textrm{C}}\bold{m}_{\mu}$. Note that $m_{\textrm{tot}}^z \equiv 1/2$ for all interaction strength, showing the ground state is ferromagnetic with a total magnetization $\mathscr{N}/2$, where $\mathscr{N}$ is the number of unit cells. These findings are consistent with Lieb's rigorous results~\cite{Lieb-theorem}. Another degenerate ground state with magnetization $-1/2$ per unit cell is also found as expected, which is physically equivalent to the one with positive magnetization shown in Fig.~\ref{fig.SU2}(a). Therefore, it suffices to focus on the one with positive magnetization in the following discussion. 

Although the total magnetization remains a constant for all $U/t$, the contributions from different sublattices vary with interaction, as shown by the magnetizations in Fig.~\ref{fig.SU2}(a) and the particle occupations $\langle n_{\mu,\alpha} \rangle$ in Fig.~\ref{fig.SU2}(b). In the weakly interacting limit, the contribution is mainly from the B and C sites. This can be understood by noting that in the non-interacting case, the flat band which is responsible for ferromagnetism is consisting of localized states on sublattices B and C $\psi (\vecc{k}) \propto (0, -\cos(k_x), \cos(k_y))^{T}$. At large $U$ limit, we find $ m_{\textrm{A}}^z  =-1/2 $, $ m_{\textrm{B}}^z =m_{\textrm{C}}^z  =1/2 $, and $\langle n_{\textrm{A},\downarrow} \rangle=\langle n_{\textrm{B},\uparrow} \rangle=\langle n_{\textrm{C},\uparrow} \rangle=1$. Namely, site A is occupied by one spin-down particle, while sites B and C are both occupied by one spin-up particle, as schematically depicted in Fig.~\ref{fig.SU2}(d). Such a staggered ferromagnetic ordering resembles the Mott insulator phase, as one would naturally expect for large repulsive interaction.

Moreover, the ferromagnetic ground state can be identified from spin-density mean fields, i.e., the number of particles with spin $\alpha$ in one unit cell, defined as $\langle n_{\alpha} \rangle=\sum_{\mu=\textrm{A,B,C}} \langle n_{\mu,\alpha} \rangle$.
Although $\langle n_{\mu,\alpha}\rangle$ varies with $U/t$ in Fig.~\ref{fig.SU2}(b), the total number of particles of each spin in one unit cell $\langle n_{\alpha}\rangle$ is independent of $U/t$. Furthermore, we find  $\langle n_{\uparrow} \rangle =2 \langle n_{\downarrow} \rangle$ for any $U/t >0$. Namely, the number of spin-up particles is twice the number of spin-down particles in every unit cell. This is another implication of the ferromagnetic state. 

We further analyze the correlation functions defined as~\cite{Shen}
\begin{eqnarray}
m(0) &=& \frac{1}{\mathscr{N}} \big\langle \big(\sum_{i}S_{i}^{+}\big)  \big(\sum_{j}S_{j}^{-}\big) \big\rangle,
\nonumber \\
m(Q) &=&\frac{1}{\mathscr{N}} \big\langle \big(\sum_{i} \epsilon (i)S_{i}^{+}\big)  \big(\sum_{j}\epsilon (j)S_{j}^{-}\big) \big\rangle,
\end{eqnarray} 
where the spin raising and lowering operators are defined as $S_i^{+}=c_{i,\uparrow}^{\dag}c_{i,\downarrow}$ and $S_i^{-}=c_{i,\downarrow}^{\dag}c_{i,\uparrow}$, respectively, $\epsilon(i)=1$ when site $i$ belongs to sublattice A, and $\epsilon(i)=-1$ when site $i$ belongs to sublattices B and C. Results of $m(0)$ and $m(Q)$ in the mean-field ground state as functions of $U/t$ are shown in Fig.~\ref{fig.SU2}(c), showing respectively the existence of ferromagnetic long-range order and the staggered features therein. Notice that the relation $m(Q)\ge m(0)$ holds in our mean field calculation, which is consistent with the rigorous theorem~\cite{Shen}.

\begin{figure}
\centering
\includegraphics[width=1.0\columnwidth]{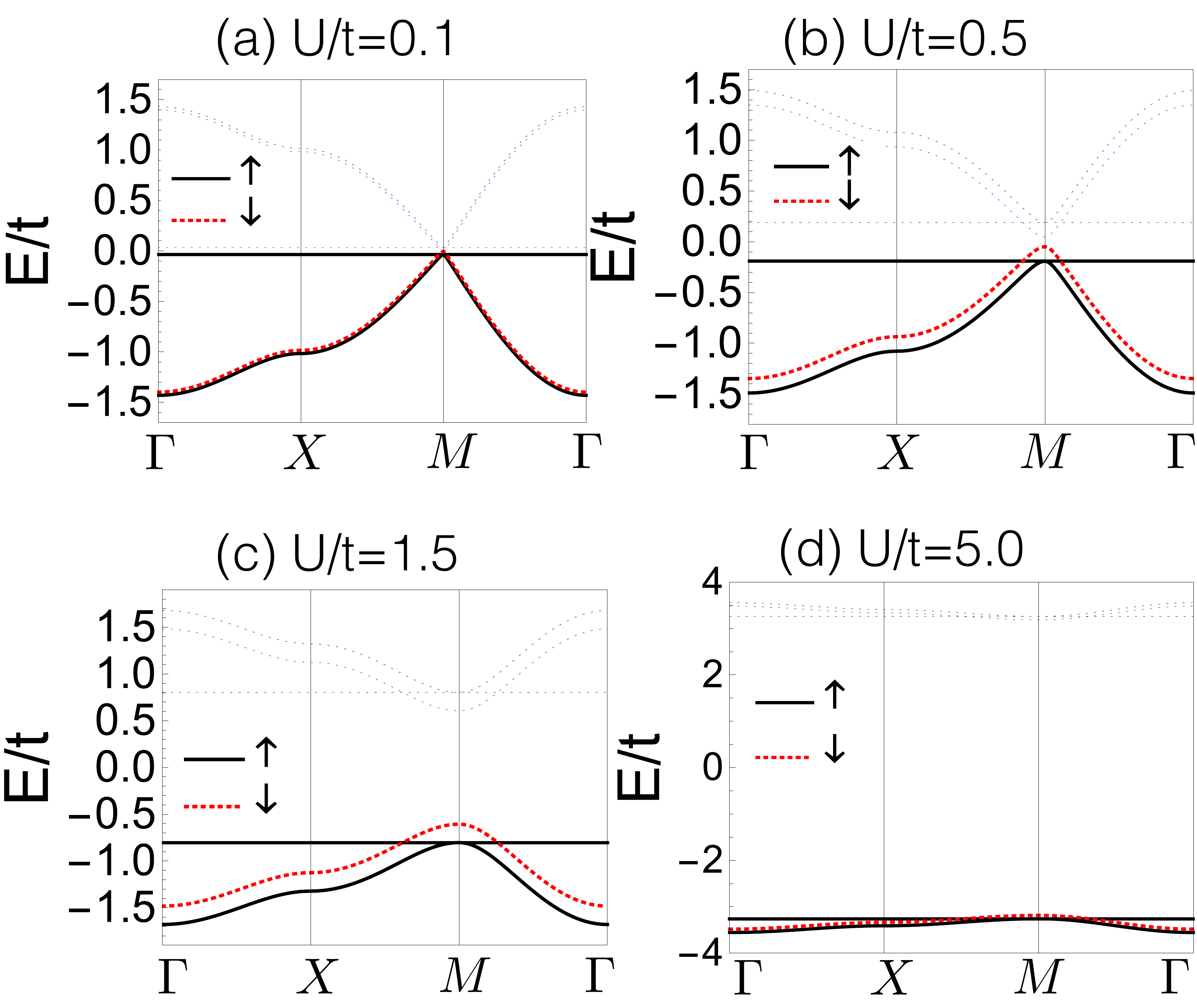}
\caption{The zero-temperature band structure of the SU(2) model at half filling with different interactions: (a) $U/t=0.1$, (b) $U/t=0.5$, (c) $U/t=1.5$ and (d) $U/t=5.0$. The band filled by spin-up (down) particles is labeled by thick black solid (red dashed) line. The rest three bands labeled by thin black dotted lines are empty.}
\label{fig.extendedBZ-SU2}
\end{figure}

In time-of-flight experiment, by tuning the laser power, the optical lattice potential can be dynamically controlled and adiabatically removed. During this adiabatic process, each Bloch wave is mapped to a free-particle momentum  state~\cite{Bloch-RMP}. Such band mapping technique can be used to probe particle's quasi-momentum distribution~\cite{Bloch-RMP, TOF-BEC,TOF-fermion}. To compare with the particle absorbing image in time-of-flight experiment, we plot in Fig~\ref{fig.extendedBZ-SU2} the band structure for different interacting strengths, and highlight the lowest three bands which are filled at zero temperature. We find that at half filling, the first (black solid), the second (red dashed) and the third (black solid) bands are occupied by spin-up, spin-down, and spin-up particles, respectively. The rest three empty bands are labeled by thin black dotted lines. 

\begin{figure}
\centering
\includegraphics[width=0.6\columnwidth]{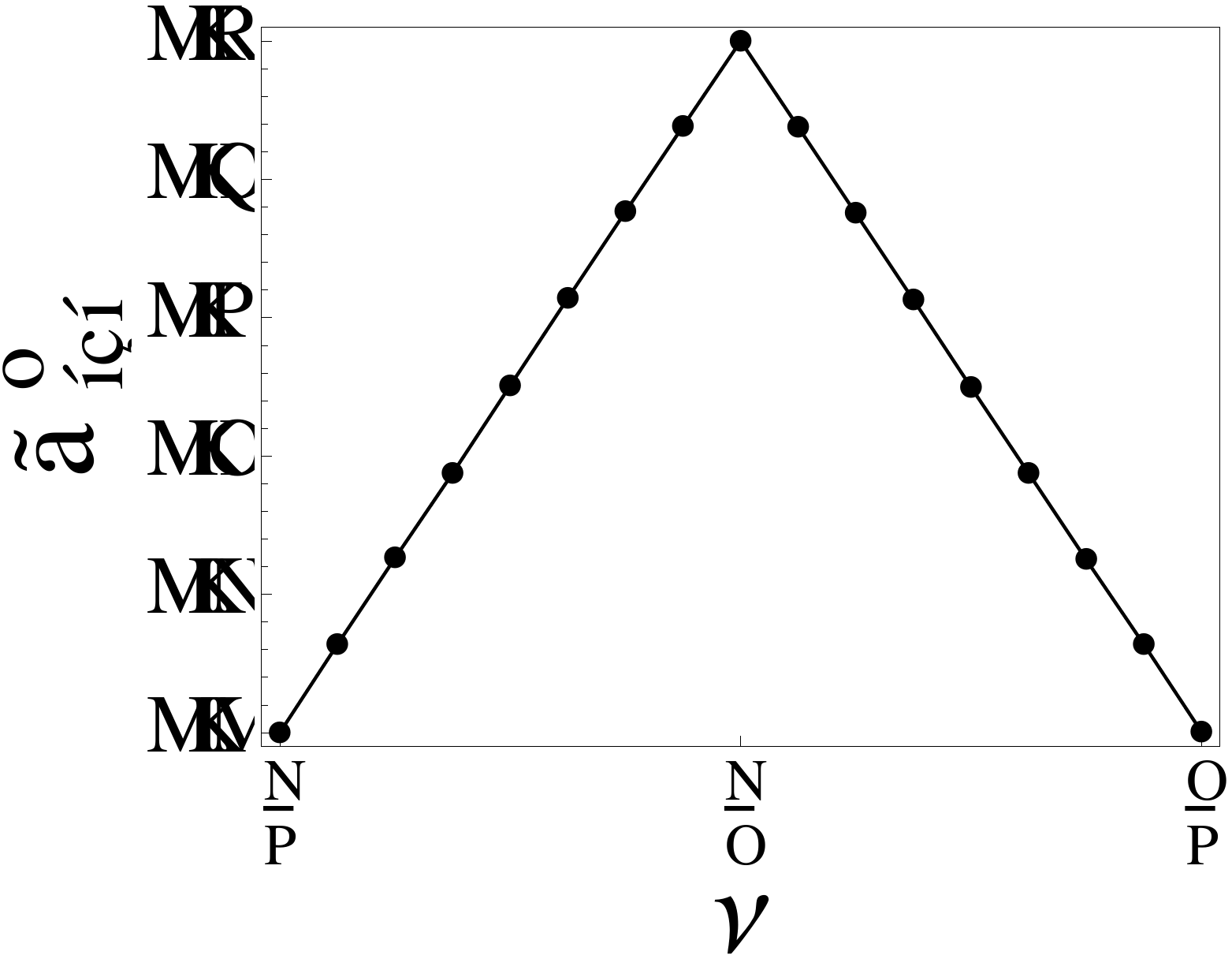}
\caption{The total magnetization in one unit cell $m^z_{\textrm{tot}}$ for the SU(2) model with $U/t=0.0505$. Similar to the half-filling case, the magnetization orders are collinear: $\bold{m}_{\textrm{B}}$ is parallel with $\bold{m}_{\textrm{C}}$ and antiparallel with $\bold{m}_{\textrm{A}}$, where $\bold{m}_{\textrm{B}}=\bold{m}_{\textrm{C}}$. Therefore, we only plot total magnetization in one unit cell $m^z_{\textrm{tot}}$ ($m^{x,y}_{\textrm{tot}}=0$ are not shown).}
\label{fig.m-filling-SU2}
\end{figure}

Next, we go beyond the half filling case and investigate the relation between magnetization and filling fraction $\nu$ within the regime $1/3 \le \nu \le 2/3$, where the Lieb's theorem does not hold any longer. In this range, the flat band is partially filled in a non-interacting system. As the flat band is crucial in Lieb's argument, we expect that the flat-band ferromagnetism still holds at least in the weak-interacting limit, and the magnetization should be proportional to the filling fraction when $1/3 \le \nu \le 2/3$ and $U/t \ll 1$. In Fig.~\ref{fig.m-filling-SU2}, we show the mean-field result for the magnetization with increasing filling fraction, and indeed observe a linear dependence. Specifically, we find that the magnetization becomes nonzero as long as the flat band starts to get filled, leading to a magnetized state throughout the entire regime of $1/3 \le \nu \le 2/3$. Such an observation seems to be qualitatively different from a previous study~\cite{percolation}, that takes Tasaki lattice~\cite{TasakiPRL,Mielke-Tasaki} as an example, maps the flat band ferromagnetism to the Pauli-correlated percolation problem and shows that the ferromagnetism can only exist when the flat band is filled to some extend. The Tasaki lattice can be regarded as the Lieb lattice with nearest-neighbor hopping among sublattice A, in which case the bipartiteness breaks down. As a result, the lowest band is tuned to be flat. We emphasize that the method of mapping the flat band ferromagnetism to the Pauli-correlated percolation, which requires the number of particles per trapping cell is no more than one~\cite{percolation}, does not apply to the Lieb lattice problem discussed here, as the number of particles per trapping cell is more than two when the flat band is partially filled. 

Finally, we consider the residual symmetry of the ferromagnetic phase of the SU($2$) Hubbard model. When the SU($2$) symmetry breaks, resulting in a two-fold degenerate ferromagnetic ground state, the U($1$) symmetry still remains, where the charge of each spin component is conserved. The symmetry breaking is thus SU$(2)$ $\rightarrow$ U$(1)$.

To conclude, we notice the mean-field calculation works well to capture the magnetic properties of the ground state of the SU($2$) Hubbard model on the Lieb lattice. We find at half filling, the SU$(2)$ symmetry is spontaneously broken into U($1$) symmetry in the ground state, resulting in a ferromagnetic state, in which the magnetization in one unit cell is $1/2$. The total magnetization is found to be independent of interaction. Our mean-field calculation is qualitatively consistent with the conclusions obtained by the Lieb's theorem. Moreover, we find the ground state supports a staggered ferromagnetic long-range order, which is consistent with Ref.~\cite{Shen}. The picture that particles filling into the flat band contribute to magnetization works well, at least in the weak-interacting limit for the SU($2$) Hubbard model. 

\section{\label{SU3}Ferromagnetic ground state of the SU$(3)$ Hubbard model}

For the Hubbard model with the SU$(3)$ symmetry, there are three species of particles, hence we take $\alpha=1,2,3$ to label the flavors or colors of the particles. As the ferromagnetic state may correspond to particles partially filled into the flat band in the non-interacting case, here we consider a typical example of $4/9$ or $5/9$ filling, which are physical equivalent due to the particle-hole symmetry. Note that the filling fraction now is defined as $\nu=N_e/(3N_s)$ for SU($3$) system. We then take $N=3$ and $\vecc{\tau}=\vecc{\lambda}$ ($\lambda^a$ is a  Gell-Mann matrix in fundamental representation with $a=1,\cdots,8$, as shown in Appendix A) into Eq.~\eqref{eq.matHk} and solve the problem self-consistently. 

\begin{figure}
\centering
\includegraphics[width=1.0\columnwidth]{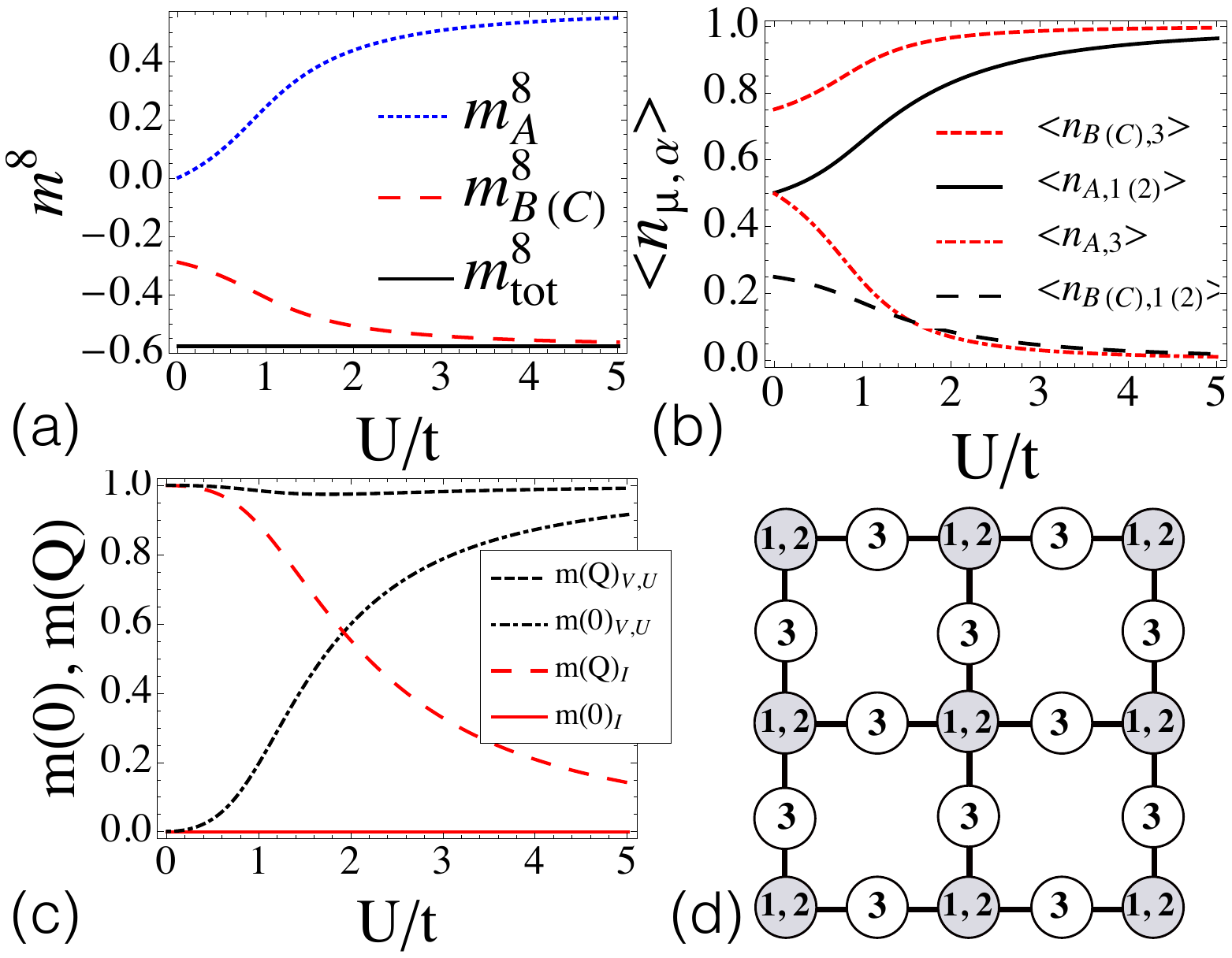}
\caption{The properties of the magnetic order for the SU(3) Hubbard model at 4/9 filling. (a) Results of $m_{\textrm{A,B(C)}}^{8}$ and $m_{\textrm{tot}}^8$ in one unit cell as functions of $U/t$. (b) The particle numbers on each site with different flavors $\langle n_{\mu,\alpha} \rangle$. (c) The correlation functions $m(0)_{I,V,U}$ and $m(Q)_{I,V,U}$ as functions of $U/t$. (d) A schematic figure showing the flavor-density distribution and  
the staggered ferromagnetic long-range order at the large-$U$ limit.}
\label{fig.SU3}
\end{figure} 

The magnetization order $\bold{m}_{\mu}$ now is an eight-dimensional vector in the spin space, which transforms according to the adjoint represent of the SU$(3)$ Lie algebra. We find that at $4/9$ filling the ground state supports a collinear magnetic order, because $\cos \theta_{<\bold{m}_{\textrm{A}},\bold{m}_{\textrm{B(C)}}>}=-1$ is found numerically. One way to investigate the collinear magnetization order in eight-dimensional space is considering the norm of total magnetization in one unit cell $|\bold{m}_{\textrm{tot}}|$. We find $|\bold{m}_{\textrm{tot}}|=1/\sqrt{3}$, which implies the ground state is ferromagnetic. Another more comprehensive way to analyze the collinear magnetization order is performing a global SU($3$) transformation to the mean-field Hamiltonian, so that the ferromagnetic ground states are denoted by $|m_{\textrm{tot}}^3, m_{\textrm{tot}}^8 \rangle$ as explained at the end of Sec.~\ref{general-MF}, where the other six components of $\bold{m_{\textrm{tot}}}$ are zero after the SU($3$) transformation. We indeed find the ferromagnetic ground state has three-fold degeneracy as expected, which can be labeled by  $| 0, -1/\sqrt{3} \rangle$ and $|\pm 1/2, 1/2\sqrt{3} \rangle$. These three degenerate states are physically equivalent  and can be transformed to each other by an SU($3$) rotation as discussed before. Therefore, it suffices to focus on any one of them. Without loss of generality, we choose the one described by $| 0, -1/\sqrt{3} \rangle$ in the following, whose magnetization in one unit cell is denoted by $(m^3=0, m^8=-1/\sqrt{3})$. 

To see how magnetization is affected by interaction and distributed on different sublattices, we plot $m_{\textrm{A,B(C)}}^{8}$  and $m_{\textrm{tot}}^8$ as functions of $U/t$ in Fig.~\ref{fig.SU3}(a). The components associated with $m_{\mu=\textrm{A,B,C}}^3$ are zero in the ground state and are not shown. 
The $C_4$ symmetry is preserved for any finite $U$ at $4/9$ filling due to $m_{\textrm{C}}^8=m_{\textrm{B}}^8$. 
Similar to the SU$(2)$ case, the ground state of the SU$(3)$ Hubbard model is ferromagnetic with a collinear magnetic orders for any finite $U/t>0$ at such filling. And the values of magnetization $m_{\rm tot}^3$ and $m_{\rm tot}^8$ are independent of $U/t$. Namely, the ferromagnetic ground state at weak-interacting limit is adiabatically connected to the one at strong-interacting limit. The ferromagnetic state implies the spontaneous SU($3$) symmetry breaking in the presence of finite $U$. 

We plot the numbers of particles with different flavors at each site $\langle n_{\mu,\alpha} \rangle$ as functions of $U/t$ in Fig.~\ref{fig.SU3}(b) to show the contribution to magnetization from different sublattices. We find the number of particles with flavor $1$ is the same as that with flavor $2$ on each sublattice, $\langle n_{\mu,1} \rangle=\langle n_{\mu,2} \rangle$. It is a consequence of residual symmetry created by SU($3$) symmetry breaking and will be discussed later. Similar to the SU($2$) systems, at the weak-interacting limit, the contribution to magnetization mainly comes from particles residing on sublattices B and C. But at large $U$ limit, we find $m_{\textrm{A}}^8=\sqrt{3}/3$, $m_{\textrm{B(C)}}^8=-\sqrt{3}/3$ and $n_{\textrm{A},1(2)}=1$, $n_{\textrm{B(C)},3}=1$. Namely, site A is occupied by two particles with flavor-$1$ and flavor-$2$ , while sites B and C are occupied by one flavor-$3$ particle, respectively. 
A schematic figure to illustrate the flavor-density distribution pattern of the SU($3$) Hubbard model at $4/9$ filling on the Lieb lattice is shown in Fig.~\ref{fig.SU3}(d).
In the intermediate interaction range, the flavor-density pattern can be summarized as: particles with two flavors (flavor-$1$ and $2$) prefer one sublattice (A) with equal density, while the particles with the third flavor (flavor-$3$) prefer the other sublattice (B and C actually belong to the same sublattice as $C_4$ symmetry is preserved here) with a larger density.

The property that the ground state is ferromagnetic can also be extracted by examining the imbalanced numbers of particles with different flavors from the flavor-density mean fields $\langle n_{\alpha} \rangle$. We find that $\langle n_1 \rangle=\langle n_2 \rangle=1$ and $\langle n_{3} \rangle=2$  in one unit cell. And the expectation value of particle number of each spin component in the ground state is independent of $U/t$. Such a fact implies a ferromagnetic state, which exists at any finite $U$. 

We also investigate the long-range correlations in the ground state of the SU$(3)$ Hamiltonian. Different from the SU($2$) system, we need to define three pairs of ``spin-raising'' and ``spin-lowering'' operators~\cite{Georgi-book}:
$I^{\pm} \equiv T^1 \pm i T^2=E_{\pm1,0}$, $V^{\pm} \equiv T^4 \pm i T^5=E_{\pm1/2,\pm \sqrt{3}/2}$, $U^{\pm} \equiv T^6 \pm i T^7=E_{\mp 1/2,\pm \sqrt{3}/2}$, where $T^{\alpha}=\lambda^{\alpha}/2$.
Then we define three pairs of correlation functions as
\begin{eqnarray}
m(0)_O&=&\frac{1}{\mathscr{N}} \big\langle \big(\sum_{i}O_{i}^{+}\big)  \big(\sum_{j}O_{j}^{-}\big) \big\rangle, \nonumber\\
m(Q)_O&=&\frac{1}{\mathscr{N}} \big\langle \big(\sum_{i} \epsilon (i)O_{i}^{+}\big)  \big(\sum_{j}\epsilon (j)O_{j}^{-}\big) \big\rangle,\nonumber
\end{eqnarray}
where $O=I,V,U$, $\epsilon(i)=1$ when $i\in A$, and $\epsilon(i)=-1$ when $i\in B(C)$.
The plots of $m(0)_O$ and $m(Q)_O$ in the mean-field ground state as functions of $U/t$ are shown in Fig.~\ref{fig.SU3}(c). We find that in the ferromagnetic ground state denoted by $|0,-{1}/{\sqrt{3}}\rangle$, the correlation functions $m(0)_{V,U}$ and  $m(Q)_{V,U}$ are nonzero, indicating a ferromagnetic long-range order with staggered features. Besides, we find that the relation $m(Q)_O \ge m(0)_O$ still holds in our mean-field calculation.

\begin{figure}
\centering
\includegraphics[width=0.85\columnwidth]{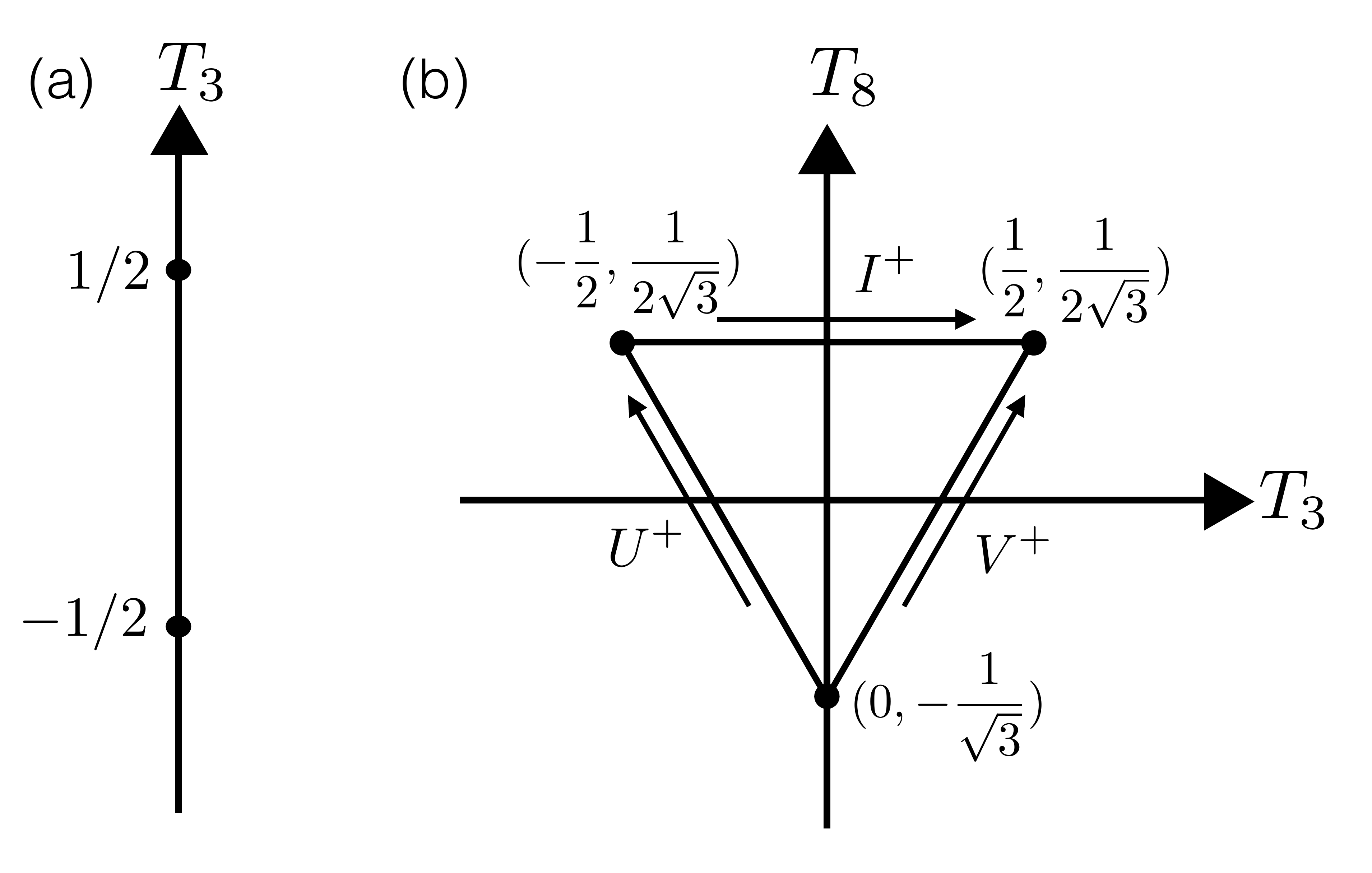}
\caption{The equivalent ferromagnetic ground states can be shown with the aid of weights diagram of (a) the SU($2$) Lie algebra and (b) the SU($3$) Lie algebra. The coordinates of the weights are labeled in each diagram.}
\label{fig.weights-diagram}
\end{figure}

Now we discuss the residual symmetry in the ground state. The ground state is a three-fold degenerate ferromagnetic state in which the SU$(3)$ symmetry spontaneously breaks to a remaining SU$(2)$$\times$U$(1)$ symmetry. One could understand this with the help of weights diagram of the SU$(3)$ Lie algebra [see Fig.~\ref{fig.weights-diagram}(b)].
For example, if $|0,-1/\sqrt{3}\rangle$ is chosen as the mean-field ferromagnetic ground state, one could see that~\cite{Georgi-book}
  \begin{eqnarray}
    I^{+}|0,-\frac{1}{\sqrt{3}}\rangle&=&0,\nonumber\\
    I^- |0,-\frac{1}{\sqrt{3}}\rangle&=&0.\nonumber
  \end{eqnarray} 
This relation shows that there is an SU$(2)$ symmetry with generators $\{T^1,T^2,T^3\}$ which leaves the mean-field ground state invariant. The extra U$(1)$ symmetry is the usual charge conservation symmetry. 
  In the general SU$(N)$ case, there would be $N$ equivalent ferromagnetic states in which the SU$(N)$ symmetry could spontaneously break and the symmetry breaking pattern would be SU$(N)$ $\rightarrow$ SU$(N-1)\times$U$(1)$.

\begin{figure}
\centering
\includegraphics[width=1.0\columnwidth]{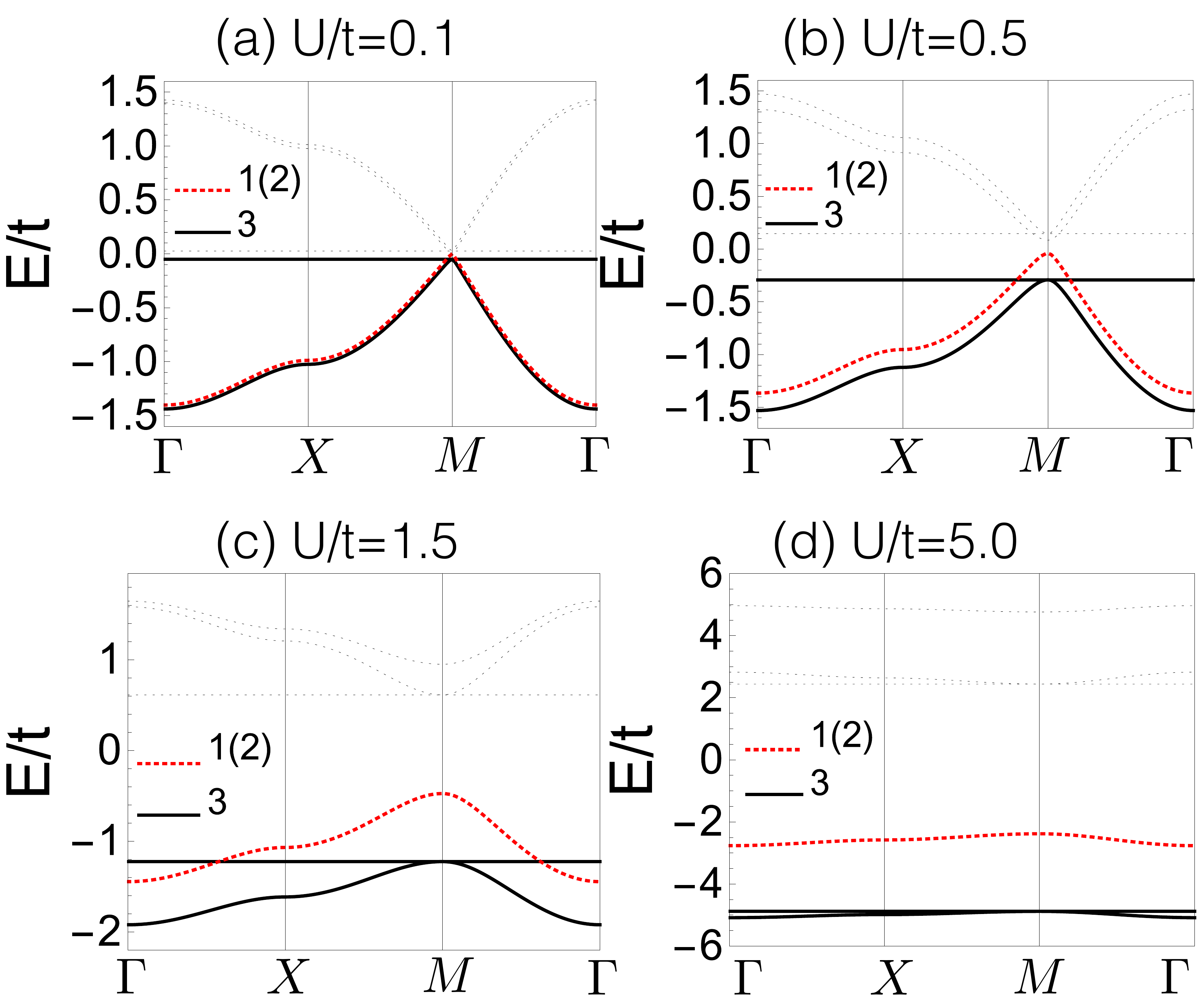}
\caption{The zero-temperature band structure of the SU(3) model at 4/9 filling with different interactions (a) $U/t=0.1$, (b) $U/t=0.5$, (c) $U/t=1.5$ and (d) $U/t=5.0$. The bands filled by flavor-$1$ or flavor-$2$ particles are labeled by red dashed lines, while the bands filled by flavor-$3$ particles are labeled by black solid lines. The empty bands are labeled by thin black dotted lines.}
\label{fig.BZ-SU3-filled}
\end{figure}

Furthermore, to compare with the absorbing image in time-of-flight experiment, we plot the band structure with different interactions as shown in Fig~\ref{fig.BZ-SU3-filled}: (a) $U/t=0.1$, (b) $U/t=0.5$, (c) $U/t=1.5$, and (d) $U/t=5.0$. We find the bands occupied by flavor-$1$ and flavor-$2$ are doubly degenerate, which are labeled by red dashed lines. The two bands occupied by flavor-$3$ particles are labeled by black solid lines, while the other five empty bands are labeled by black dotted lines.
With weak interaction as shown in Figs.~\ref{fig.BZ-SU3-filled}(a) and \ref{fig.BZ-SU3-filled}(b), the first and the forth bands are occupied by flavor-$3$ particles; the second and third bands (degenerate) are occupied by flavor-$1$ and flavor-$2$ particles, respectively. 
This observation indicates that in this weak-interacting case, one can still rely on the particle-filling picture in which the particles residing on the flat band contribute to the magnetization.

However, at large-$U$ limit as shown in Fig.~\ref{fig.BZ-SU3-filled}(d), we find the lowest two bands are occupied by flavor-$3$ particles, and the particles with flavor-$1$ and flavor-$2$ occupy two higher bands. This is because at large-$U$ limit, particles with flavor-$1$ and flavor-$2$ residing on the same sublattice A as shown in Fig.~\ref{fig.SU3}(d) obtain a much larger interaction energy comparing with kinetic energy. Therefore, the particles with flavor-$1$ and flavor-$2$ can populate into higher bands, due to the higher energy gained by the interaction. We find the flavor distribution in quasi-momentum space is sensitive to interactions. This feature can by observed in time-of-flight expansion by an adiabatic removal of the trapping potential~\cite{TOF-fermion}.

In addition, we find that the system is an insulator with a finite gap in the ground state. The presence of such a gap is allowed by generalized Lieb-Schultz-Mattis (LSM) theorem~\cite{LSM-Oshikawa}. The original LSM theorem proves the excitation gap in antiferromagnetic spin chain~\cite{LSM}. It is later extended to a quantum many-particle system (bosons or fermions) with translational symmetry and conserved particle number, where a finite excitation gap is proved to be possible only if the number of particles per unit cell is an integer in the ground state~\cite{LSM-Oshikawa}. In our settings, the particle number per unit cell is an integer, and we find the particle number of each flavor component in the ground state $\langle n_{\alpha} \rangle$ is an integer as well. Therefore, the gap found in our SU($3$) system is consistent with the generalized LSM theorem. 

\begin{figure}
\centering
\includegraphics[width=1.0\columnwidth]{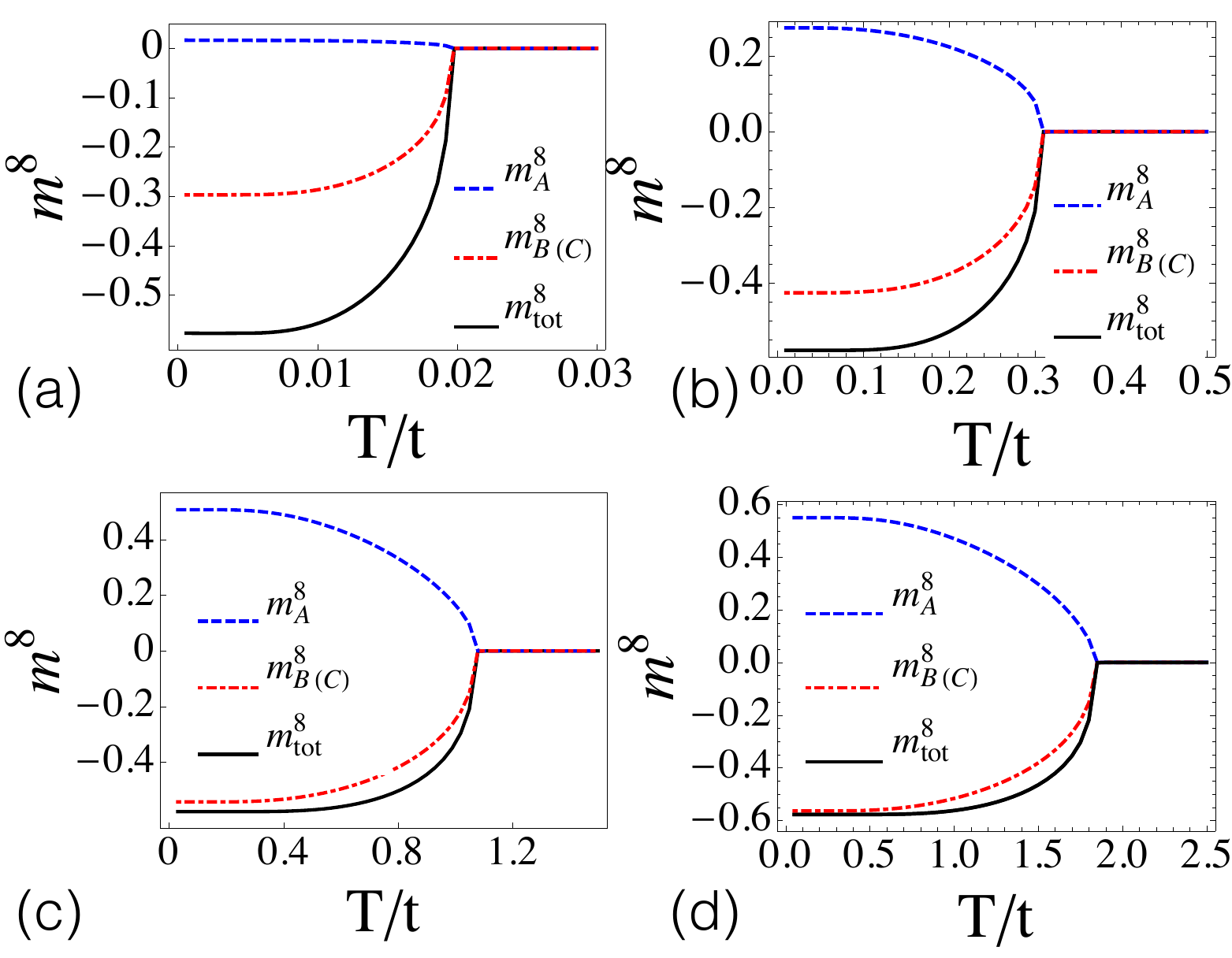}
\caption{The magnetization for the SU(3) model at finite temperature with different interactions. (a) The critical temperature $T_c/t=0.0192$ and the critical entropy per particle $S_c=0.457$ when $U/t=0.11$; (b) $T_c/t=0.31$ and $S_c=0.818$ when $U/t=1.11$; (c) $T_c/t=1.08$ and $S_c=1.406$ when $U/t=3.01$; (d) $T_c/t=1.80$ and $S_c=1.452$ when $U/t=5.01$. }
\label{fig.m_T_SU3}
\end{figure}

Moreover, we investigate the finite temperature effect on the magnetization orders with the mean-field approach. Although the Mermin-Wagner theorem forbids the existence of any long-range order with short-range interaction in two-dimensional system at finite temperatures~\cite{MerminWagner1966,MerminWagner1968}, quasi-long-range features such as magnetic domains can still be realized and observed in experiments for temperatures well below the mean-field critical temperature. 
In Fig.~\ref{fig.m_T_SU3}, we plot the magnetization as a function of temperature with different values of interaction. One can easily find that the critical temperature $T_c$ increases with interaction. For example, $T_c/t$ is around $0.0192$ when $U/t=0.11$ as shown in Fig.~\ref{fig.m_T_SU3}(a), while $T_c/t$ increases up to $1.8$ when $U/t$ is $5.01$ as shown in Fig.~\ref{fig.m_T_SU3}(d). The critical temperature would increase further with even stronger interactions. 

To make connection with experiments on cold atoms loaded in optical lattices, where the atoms are not in contact with a bath and the temperature is not a conserved quantity, we also calculate the critical entropy per particle for the ferromagnetic state as it remains unchanged when going to different interactions. The entropy is given by 
\begin{equation}
S=-k_B \sum_{i} [ f(\epsilon_i) \ln f(\epsilon_i) + (1-f(\epsilon_i)) \ln (1-f(\epsilon_i))],
\end{equation}
where $f(\epsilon_i)$ is the occupation number for a Fermi gas $f(\epsilon_i)=[\exp{((\epsilon_i-\mu)/k_B T)}+1]^{-1}$. We find the critical entropy per particle needed is $S_c=0.457$ when $U/t=0.11$,  and $S_c=1.452$ when $U/t=5.01$. Both values are significantly larger than the lowest entropy per particle that can be achieved in experiment ($\sim 0.2$). Therefore, in experiment the ferromagnetic ground state could be observed in cold atomic systems loaded into an optical lattice with increasing density-density on-site interaction.

\begin{figure}
\centering
\includegraphics[width=1.0\columnwidth]{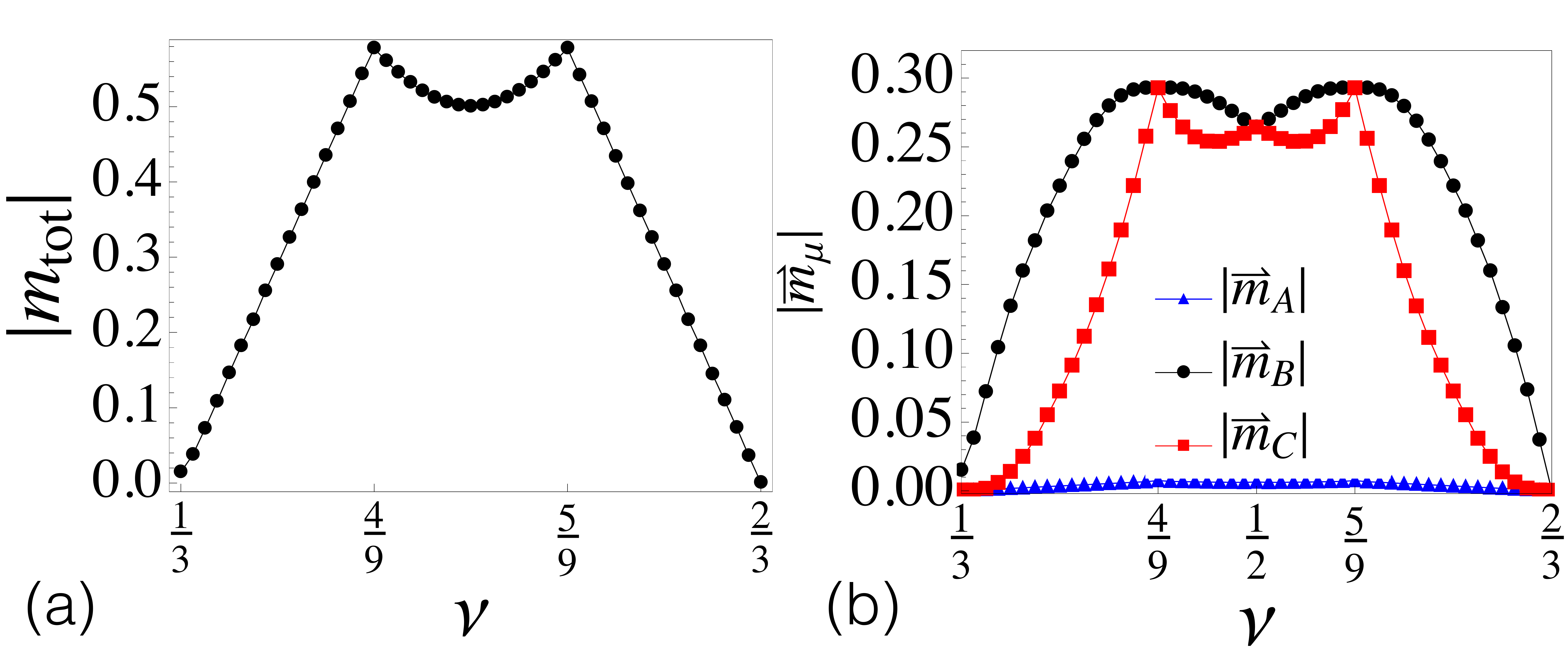}
\caption{(a) The norm of the total magnetization in one unit cell $|\bold{m}_{\textrm{tot}}|$ for the SU(3) model; (b) the norms of magnetization on each sublattice site $|\bold{m}_{\mu}|$ for the SU(3) model, where $\mu=\textrm{A,B,C}$. The interaction strength used in this plot is within the weakling interacting limit with $U/t=0.0505$. }
\label{fig.m_filling_SU3}
\end{figure}

The discussion above focuses on a fixed filling factor of $4/9$, which corresponds to the case that the flat band is one-thirdly filled. Next, we extend our calculation to other filling fractions $\nu$ within the regime $1/3<\nu<2/3$, which corresponds to the case that the flat band is partially filled. In Fig.~\ref{fig.m_filling_SU3}(a), we show the total magnetization within one unit cell $|\bold{m}_{\textrm{tot}}|$ by varying filling faction for a weak interaction with $U/t=0.0505$. Note that the magnetization reaches a maximum at $\nu=4/9$ and $5/9$, as one would naturally expect in the weak-interacting limit. Furthermore, we find $|\bold{m}_{\textrm{tot}}|$ increases almost linearly with filling fraction $\nu$ when $1/3<\nu<4/9$, and decreases with an opposite trend with $\nu$ when $5/9<\nu<2/3$, as required by the particle-hole symmetry. In these two filling regimes, the magnetization vectors on the B and C sublattices $\bold{m}_{\textrm{B}}$ and  $\bold{m}_{\textrm{C}}$ keep antiparallel with that on the A sublattice $\bold{m}_{\textrm{A}}$. However, in the filling regime $4/9<\nu<5/9$, the magnetization vectors $\bold{m}_{\textrm{B}}$ and $\bold{m}_{\textrm{C}}$ are neither parallel nor antiparallel with $\bold{m}_{\textrm{A}}$, resulting in a non-collinear magnetic order. Besides, we also find that the magnetization on the B and C sublattices are not equivalent as $\nu$ deviates from $4/9$ and $5/9$, as shown in Fig.~\ref{fig.m_filling_SU3}(b). Such a spontaneous breaking of $C_4$ rotational symmetry thus implying a nematic ferromagnetic state~\cite{Zhang,Tsai}. 

\begin{figure}
\centering
\includegraphics[width=0.7\columnwidth]{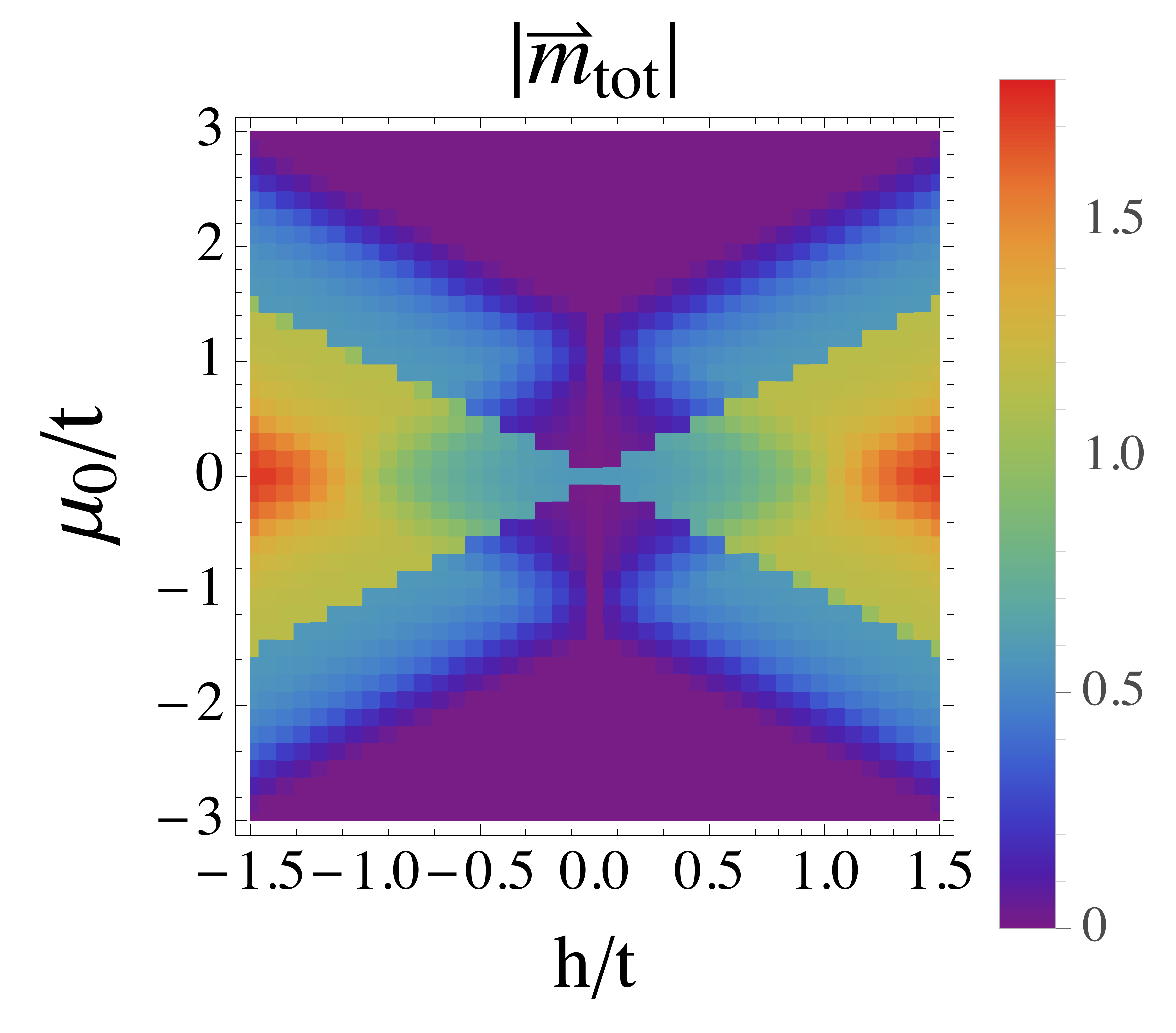}
\caption{The norm of total magnetization for the SU(3) model in one unit cell $|\bold{m}_{\textrm{tot}}|$ in the plane of $(h/t, \mu_0/t)$ with $U/t=0.0505$. }
\label{fig.phasediagram}
\end{figure}

To make connections to experiments where an inhomogeneous global trapping potential $V_{\rm ext}({\bf r})$ is usually present, we further consider the case of grand canonical ensemble
\begin{equation}
\mathcal H=H_0 - \mu_1 \sum_{i} n_{i,1} -\mu_2 \sum_{i} n_{i,2} - \mu_3 \sum_{i} n_{i,3},
\label{eq.newHam}
\end{equation}
where $\mu_{\alpha=1,2,3}$ are chemical potentials of atoms of spin $\alpha$. Within the local density approximation, the phase one would encounter by traversing the global trap is then determined by the position dependent chemical potentials $\mu_{\alpha} ({\bf r}) = \mu_{\alpha}({\bf r} = 0) - V_{\rm ext}({\bf r})$. As a typical example, we show in Fig.~\ref{fig.phasediagram} the total magnetization within a unit cell by varying $\mu_0=(\mu_1+\mu_3)/2$ and $h=(\mu_1-\mu_3)/2$, with the condition $\mu_1=\mu_2$.
Notice that the ferromagnetic state appears in a wide non-purple regime on this phase diagram, which would facilitate the realization and observation of such a state in experiments. Besides, we find $|\bold{m}_{\textrm{tot}}|=0.57735=\sqrt{3}/3$ at the point $(h/t=0,\mu_0/t=0)$, which is consistent with our calculation for Eq.~(\ref{eq.Ham}). Meanwhile, by increasing $|h/t|$ with fixed $\mu_0=0$, the magnetization $|\bold{m}_{\textrm{tot}}|$ increases with $|h/t|$ and finally saturates to $1.73205=\sqrt{3}$ around $|h/t| = 1.5$. This is a natural result as the chemical potential difference $h=(\mu_1-\mu_3)/2\ne0$ is equivalent to an effective magnetic field which can introduce polarization. Specifically, we find $n_1=n_2=3$ and $n_3=0$ at the point of $(h/t=1.5, \mu_0/t=0)$, while  $n_1=n_2=0$, $n_3=3$ at the point of $(h/t=-1.5, \mu_0/t=0)$, both representing a fully polarized ferromagnetic state. 


\section{\label{conclusion}Conclusions and discussions}
In conclusion, we study at the mean-field level the SU($2$) and SU($3$) repulsive Hubbard models on the Lieb lattice, focusing on the existence and properties of a ferromagnetic state when the flat band is partially filled. For the SU($2$) case, we find that the mean-field calculation works well to capture the essential physics of the magnetic ground state. Specifically, the mean-field approach concludes that the ground state is ferromagnetic for any finite repulsive interaction, in consistent with known rigorous results. We then extend the mean-field calculation to the case of SU($3$) Hubbard model, and find that the ground state is also a ferromagnetic state as the flat band is partially filled with filling factor $\nu \in (1/3, 2/3)$. Such a magnetic state is staggered within a unit cell, where the magnetizations on the composite sublattices are different in both magnitude and direction. We also find that the $C_4$ rotational symmetry of the Lieb lattice would spontaneously break for filling away from $4/9$ and $5/9$, leading to a nematic magnetic order. To make connection to experimental conditions, we further calculate the mean-field critical temperature of the magnetic state. We find that the critical temperature increases with interactions, and the critical entropy per particle needed is much larger than the lowest one can be achieved in experiment. Therefore, our discoveries could be observed at an experimentally achievable temperature by increasing the density-density interaction. 

Finally, we give some discussion about possible competing phases. Previous studies of the SU($N$) Hubbard model on the square lattice show that for large values of $N$ (e.g. $N>6$), the ground state supports a staggered flux phase which breaks time-reversal as well as lattice translational symmetries but does not break SU($N$) symmetry~\cite{Affleck1,Affleck2,Honerkamp1}. For $N<6$, the dominant tendency of the Hubbard model at half filling is to break SU($N$) symmetry and lattice translational symmetry, resulting in a flavor density wave state~\cite{Honerkamp1}. Compared with these existing works, the present paper focuses on a special type of lattice called the Lieb lattice, which features a completely flat band in the middle of the band structure, and can support a so-called flat band ferromagnetic state at half filling for SU($2$) Hubbard model. Namely, there is no other competing phase at this special filling. Although a rigorous proof for the uniqueness of the ground state is not available so far for SU($3$) systems, the mean-field calculation presented above suggests that the ground state is also a ferromagnetic state, at least around the special filling factors of $4/9$ and $5/9$. When doping away from these special fillings, however, it is possible to have other competing phases and a much richer phase diagram, especially when the Fermi surface lies within the dispersive bands.



\section{Acknowledgement}
We thank helpful discussion with H. Zhai, M. Oshikawa, Z.-X. Liu, W. Chen, W. Zheng, L.-Y. Wang, R.-Q. He and Z.-X. Li. W.N. is supported by NSFC (11704267) and in part by the MOST of China under Grant No. 2017YFB0405700. W.Z. is supported by NSFC (11274009, 11434011, 11522436), NKBRP (2013CB922000) and the Research Funds of Renmin University of China (10XNL016, 16XNLQ03).

\appendix
\section{Gell-Mann matrices}
The Gell-Mann matrices we used in three-dimensional fundamental representation are:
\[\lambda^1=\left( \begin{array}{ccc}0 & 1 & 0\\1 & 0 & 0\\0 & 0 & 0 \end{array} \right),
\lambda^2=\left( \begin{array}{ccc}0 & -i & 0\\i & 0 & 0\\0 & 0 & 0 \end{array} \right),
\lambda^3=\left( \begin{array}{ccc}1 & 0 & 0\\0 & -1 & 0\\0 & 0 & 0 \end{array} \right),\]
\[\lambda^4=\left( \begin{array}{ccc}0 & 0 & 1\\0 & 0 & 0\\1 & 0 & 0 \end{array} \right),
\lambda^5=\left( \begin{array}{ccc}0 & 0 & -i\\0 & 0 & 0\\i & 0 & 0 \end{array} \right),
\lambda^6=\left( \begin{array}{ccc}0 & 0 & 0\\0 & 0 & 1\\0 & 1 & 0 \end{array} \right),\]
\[\lambda^7=\left( \begin{array}{ccc}0 & 0 & 0\\0 & 0 & -i\\0 & i & 0 \end{array} \right),
\lambda^8=\frac{1}{\sqrt{3}}\left( \begin{array}{ccc}1 & 0 & 0\\0 & 1 & 0\\0 & 0 & -2 \end{array} \right).\]

\section{Technical part of the general mean-field approach for the SU($N$) Hubbard model}
In the presence of a repulsive interaction, the Hubbard model on the two-dimensional Lieb lattice can not be exactly solved. And we resort to mean-field approximation. We rewrite the density-density interaction in spin channel first. Considering the traceless property of the SU($N$) generators $\textrm{Tr}\; T^a=0$ and $\textrm{Tr} (T^a T^b)=\delta_{ab}/2$, we obtain the completeness relation of the generators: 
\begin{equation}
\sum_{a=1}^{N^2-1} T^a_{\alpha\beta} T^a_{\mu\nu}
 =\frac{1}{2}\delta_{\alpha\nu}\delta_{\beta\mu}-\frac{1}{2N}\delta_{\alpha\beta}\delta_{\mu\nu}.
\end{equation}
Applying the completeness relation, we obtain the following identity
\begin{equation}
\bold{S}_i^2  =(N+1)n_i/2 -(N+1)n_i^2/(2N).
\end{equation}
Therefore, the density-density interaction is rewritten in spin/flavor space as Eq.~\eqref{eq.interaction}.

The matrix $H_{\bold{k}}^{\textrm{MF}}$ in Eq.~\eqref{eq.matHk} can be diagonalized by a unitary transformation
$U(\bold{k}) H_{\bold{k}}^{\textrm{MF}} U^{\dag}(\bold{k}) = \Lambda_{\bold{k}}$, 
where $\Lambda_{\bold{k}}$ is a diagonal matrix,  $\Phi_{\bold{k}} = U(\bold{k})\Gamma_{\bold{k}}$, and 
$\Gamma_{\bold{k}}= (\gamma_1(\bold{k}),\cdots,\gamma_{3N}(\bold{k}))^{T}$.
We then obtain $3N$ energy bands with eigenvalues $E_m (\bold{k})$ $(m=1,\cdots,3N)$, and corresponding eigenfunctions.
For a fixed filling fraction, the wavefunction at zero temperature is defined as 
\begin{equation}
| \textrm{GS} \rangle = \prod_{\bold{k}\in\textrm{BZ}} \prod_{m|E_m(\bold{k})\le E_F} \gamma_m^{\dag} (\bold{k}) |0\rangle.
\label{eq.GS}
\end{equation}
In the self-consistent mean-field calculation, we need to calculate the magnetization at each iteration step, which is defined as the expectation value of spin operators in the ground state: 
\begin{eqnarray}
m_{\mu}^a &=& \langle S_{\mu}^a (\bold{R}) \rangle =\sum_{\alpha\beta} \langle c_{\mu,\alpha}^{\dag}(\bold{R}) \frac{\tau_{\alpha\beta}^a}{2}  c_{\mu,\beta}(\bold{R})\rangle  \nonumber\\
&=&\frac{1}{2\mathscr{N}}\sum_{\bold{k}\in\textrm{BZ}}\sum_{\alpha\beta}
\langle c_{\mu,\alpha}^{\dag}(\bold{k})\tau_{\alpha\beta}^a  c_{\mu,\beta}(\bold{k})\rangle,
\label{eq.self-consistent-eq}
\end{eqnarray}
where the translational symmetry is assumed. Equation~\eqref{eq.self-consistent-eq} is the self-consistent equation for the mean-field calculation. We can write the matrix form of the unitary transformation of $U(\bold{k})$,
$c_{\mu,\alpha} (\bold{k})=\sum_m U_{\mu,\alpha,m} (\bold{k}) \gamma_m (\bold{k})$.
With the definition of the ground state in Eq.~\eqref{eq.GS}, we find that
\begin{eqnarray}
&&\langle  c_{\mu,\alpha}^{\dag} (\bold{k}) c_{\mu,\beta} (\bold{k})\rangle_{\textrm{GS}} \nonumber \\
&=&\sum_{mn} U_{\mu,\alpha,m}^{*} (\bold{k}) U_{\mu,\beta,n}(\bold{k})  
\langle \textrm{GS} | \gamma_m^{\dag}(\bold{k}) \gamma_n (\bold{k}) | \textrm{GS} \rangle \nonumber \\
& =&\sum_{m|E_m(\bold{k})\le E_F} U_{\mu,\alpha,m}^{*}(\bold{k}) U_{\mu,\beta,m}(\bold{k}).
\end{eqnarray}

\bibliography{biblio}

\end{document}